\documentclass[aps,prl,twocolumn,a4paper,10pt, longbibliography]{revtex4-1}

\usepackage[T1]{fontenc}		
\usepackage[english]{babel}		
\usepackage[latin1]{inputenc}	
\usepackage{times}

\usepackage{amsmath}			
\usepackage{amssymb}			
\usepackage{bbm}				
\usepackage{mathtools}
\usepackage{gensymb}
\usepackage{simpler-wick}
\usetikzlibrary{tikzmark}
\DeclarePairedDelimiterX\Dirbraket[3]{\langle}{\rangle}%
{#1\,\delimsize\vert\,\mathopen{}#2\,\delimsize\vert\,\mathopen{}#3}

\usepackage[colorlinks=true, a4paper=true, pdfstartview=FitV,linkcolor=blue, citecolor=blue, urlcolor=blue]{hyperref}

\usepackage{graphicx}			
\usepackage{graphics}			
\DeclareGraphicsExtensions{.pdf}
\usepackage{color}		

\newcommand{\bea}{\begin{eqnarray}}
\newcommand{\eea}{\end{eqnarray}}

\newcommand{\bk}{\mathbf{k}}

\newcommand{\Lagr}{\mathcal{L}}

\begin{document}

\title{Hund-projected Kanamori model: an effective description of Hund's metals near the Mott insulating regime}

\author{Johan~Carlstr\"om }
\affiliation{Department of Physics, KTH--Royal Institute of Technology, SE-10691 Stockholm, Sweden}
\date{\today}

\begin{abstract}
Hund's coupling plays a decisive role in shaping electron correlations of multi-orbital systems, giving rise to a class of materials--Hund's metals--that combine local-moment physics with metallic transport. Here we derive an effective low-energy description of such a system near the Mott insulating regime, starting from the multi-orbital Hubbard-Kanamori Hamiltonian and projecting onto the high-spin manifold favored by Hund's first rule. The resulting Hund-projected Kanamori model captures the interplay between carrier motion and magnetic correlations in the presence of strong Hund's coupling. In the undoped limit, the model reduces to a spin-$N/2$ Heisenberg system with suppressed quantum fluctuations, approaching the classical limit for realistic five-band configurations. Upon doping, carrier motion couples strongly to the spin background and drives ferromagnetic correlations through a Hund-enhanced kinetic mechanism analogous to, but much stronger than, Nagaoka ferromagnetism. Owing to its reduced sign problem, the model can be addressed with advanced path-integral methods to determine quasiparticle structure and effective interactions between carriers-quantities that are challenging to obtain with other methods. This framework establishes a microscopic bridge between the Kanamori model and the emergent magnetic and transport phenomena characteristic of Hund's metals.
\end{abstract}
\maketitle

Strongly correlated electronic systems occupy a central position in condensed matter physics owing to their propensity to host emergent quantum phases such as unconventional superconductivity, spin and charge order, orbital-selective states, and non-Fermi-liquid behavior. For decades, the study of strong correlations was almost synonymous with high-temperature superconductivity and microscopic descriptions based on doped Mott insulators \cite{RVB, RevModPhys.78.17,RevModPhys.84.1383}. More recently, correlation-driven phenomena have also been realized in moiré heterostructures, where a small twist angle or layers with an incommensurate interatomic spacing  produces superlattices with anomalously narrow bands and enhanced electronic interactions \cite{Cao2018,Cao2018SC,Zhang2020,PhysRevResearch.2.033087,PhysRevLett.121.026402,Xu2022}.

A central theme uniting these systems is the competition between magnetic order and carrier delocalization \cite{brink,PhysRevB.39.6880,Koepsell2019,doi:10.1126/science.abe7165}. This interplay is believed to underlie a variety of emergent states, including high-temperature superconductivity \cite{doi:10.1126/science.235.4793.1196,PhysRevB.39.11663,PhysRevB.37.1597}, quantum spin-liquids \cite{Glittum2025}, and kinetic magnetism \cite{PhysRev.147.392,PhysRevLett.95.087202,PhysRevResearch.4.043126,Tao2024}, where magnetic fluctuations can mediate effective pairing interactions leading to phase-separation and condensation \cite{PhysRevB.97.140507,CML2024}. 

In conventional strongly correlated lattice descriptions such as the Hubbard or 
the
t\text{-}J models, as well as multi-orbital generalizations thereof, strong quantum fluctuations suppress magnetic order even in the absence of charge carriers, particularly in low-spin systems \cite{Auerbach1994}.
In contrast, multi-orbital compounds with strong Hund's coupling--such as the iron pnictides, chalcogenides, and ruthenates--exhibit a qualitatively different form of correlated behavior \cite{Georges_2013}. 
These so-called Hund's metals display large local magnetic momenta owing to Hund's first rule, that persist from Mott insulating states well into the metallic regime. 
The impact of Hund's coupling  on charge-fluctuations is somewhat contradictory: At half filling it increases the effective contact repulsion, driving Mott insulating behavior. Away from half filling, it enhances charge fluctuations, promoting metallicity instead--a dual behavior referred to as the Janus effect \cite{PhysRevLett.107.256401}.
The tendency to form large local spin moments should be expected to dramatically reduce quantum spin-fluctuations. At the same time, compliance with Hund's rule should clearly make delocalization especially sensitive to spin-correlations. The emerging picture is thus of competition between almost classical magnetism, and a very potent form of kinetic ferromagnetism. This clearly sets Hund's metals apart from other strongly correlated systems.

Materials with strong Hund's coupling 
 host an array of unconventional phenomena, including multiband superconductivity, orbital-selective Mott phases, and broken time-reversal symmetry \cite{Kamihara2008,deMedici2009,Maiti2015}. Recent experiments even indicate the existence of a new fermion quadrupling state in the iron pnictide Ba$_{1-x}$K$_x$Fe$_2$As$_2$ \cite{Grinenko2021}. The diversity of materials exhibiting these behaviors highlights the importance of a general theoretical framework capable of connecting Hund's coupling, magnetism, and charge motion. 
 Because many of these systems lie close to the Mott insulating regime in terms of filling \cite{PhysRevLett.112.177001,PhysRevB.96.045133,PhysRevB.84.100509}, it is natural to begin by characterizing the correlated magnetic parent state from which the metallic phase emerges and subsequently add carriers.

In this work, we derive such a framework, starting from the multi-orbital Hubbard-Kanamori Hamiltonian and projecting onto the high-spin manifold favored by Hund's rule. The resulting Hund-projected Kanamori Model (HPKM) constitutes an effective low-energy theory for doped Hund's systems near the Mott regime derived with two assumptions: That charge-fluctuations are short-lived and can be integrated out, and that fluctuations to lower-spin states--violating Hund's first rule--are small. Despite the complexity of the original Hamiltonian, the projected model takes the form of a doped spin system with strongly suppressed quantum fluctuations and a tractable structure well-suited for path-integral and diagrammatic methods. It thereby provides a microscopic foundation for studying the magnetic dynamics and carrier-induced phenomena characteristic of Hund's metals beyond mean-field theory.

\subsection{The Kanamori Model}
The paradigmatic model of Hund's metals is the Hubbard-Kanamori Hamiltonian \cite{Georges_2013}, with an on-site interaction on the form:
\bea\nonumber
H_{\text{K}}\!\!=\!U\!\sum_{\alpha}n_{\alpha\uparrow}n_{\alpha\downarrow}\!\!+\!U'\!\sum_{\alpha\not = \beta}n_{\alpha\uparrow}n_{\beta\downarrow}
\!+\!(U'\!\!-\!J\!)\!\sum_{\alpha>\beta}\!n_{\alpha\sigma}n_{\beta\sigma}\\ 
-J\sum_{\alpha\not=\beta } a_{\alpha \uparrow}^\dagger a_{\alpha \downarrow}  a_{\beta \downarrow}^\dagger a_{\beta \uparrow}-P \sum_{\alpha\not=\beta } a_{\alpha \uparrow}^\dagger a_{\alpha \downarrow}^\dagger  a_{\beta \downarrow} a_{\beta \uparrow}.
\nonumber\\
 \label{HK}
\eea
Here, $\alpha$ denotes the band index. 
The principal energy scales of this model are the Coulomb repulsion $U$, and the Hund's coupling $J$. The inter-orbital repulsion between electrons with opposite spin is given by $U'= U-2J$, while the interaction between electrons in different orbitals with the same spin is reduced further to $U'-J$. For the pair-conversion amplitude we generally have $P=J$, though we will treat this term on a separate footing. 
The most general form of the hopping term is given by:
\bea
H_0=\sum_{ij\alpha\beta\sigma\sigma'}t_{ij\alpha\beta}a_{i\alpha\sigma}^\dagger a_{j\beta\sigma'}.
\label{hopping}
\eea
For simplicity, we restrict ourselves to nearest-neighbor processes that preserve both spin and orbital quantum numbers, though the generalization to other cases is straightforward.

The principal effect of Hund's coupling is to maximize total spin, consistent with Hund's first rule. 
With $N$ bands and the constraint that the orbital occupancy is given by $\bar{o}=[o_1,o_2...o_N]$, the Kanamori Hamiltonian (\ref{HK}) predicts a ground state spanned by
\bea
|S_M^\text{n},S_z\rangle \;, \text{with } |S_z|\le S_M^\text{n},\label{GS}\;\text{where}\\
S_M^\text{n}=\frac{1}{2}n,\;\;n=\frac{1}{2}\sum_{i=1}^N [1-(-1)^{o_i}].\label{defNS}
\eea
Here, $n$ is the number of orbitals that contribute to total spin (by being singly occupied) and $S_M^n$ is the corresponding maximal spin.
 For $H_K[P=0]$, and occupation number $n$, the first excited state is given by: 
\bea
|S_M^\text{n}-1,S_z\rangle \;\text{with } E\Big(S_M^\text{n}-1\Big)=E\Big(S_M^\text{n}\Big)+n J,\label{EX1}
\eea
which defines a ``Hund's gap'' $\Delta_H=n J$, associated with violating Hund's first rule. 
For real materials, this energy scale can be quit large: Recent estimates for iron-based superconductors--see Table \ref{t1}--indicate that $J\sim 0.3-0.6$ eV, implying a gap of $\Delta_H\sim 1.5-3$ eV at half filling, which is much greater than the energy scale of any itinerant terms in the Hamiltonian. Processes that excite the system by $\Delta_H$ will thus be strongly suppressed, prompting us to introduce a Hund's projector: 
\bea
P^H=\Pi_j \sum_{\bar{o}^j} \sum_{S_z^j=-S_M^{n^j}}^{S_M^{n^j}} |S^{n^j}_{M},S^j_z\rangle \langle  S^{n^j}_{M},S^j_z|,
\label{PH}
\eea
where $\bar{o}^j$ is the orbital occupation of the site $j$ and $n_j=n_j[\bar{o}^j]$ is given by Eq. (\ref{defNS}). The operator (\ref{PH}) thus projects onto states satisfying the condition that the total spin on any given site is maximal for its given orbital occupation, consistent with the Kanamori Hamiltonian (\ref{HK}) and Hund's first rule. Eventually, we will constrain our Hilbert space to the subspace on which $P^H$ projects, while considering short-lived excitations across the Hund's gap as emergent processes.

An additional effect of Hund's coupling is its impact on charge fluctuations. Given two sites  where all orbitals are singly occupied, in a ground state of the type (\ref{GS}), the energy cost of creating a particle-hole pair is given by
\bea
U_{\text{eff}}\!=\!E(N\!+\!1)\!+\!E(N\!-\!1)\!-\!2E(N)\!=\!U\!+\!J(N\!-\!1).\label{Ue}
\eea
At half filling, a key consequence of Hund's coupling is thus a substantial increase of the effective on-site repulsion, which drives Mott insulating behavior. It should be noted however, that away from half filling, Hund's coupling instead promotes metallicity \cite{Georges_2013}.

\begin{table}[h!]
\centering	
\begin{tabular}{ cc} 
 \hline
 Term & Energy scale		\\
 $U$ & $2.2-4.2$ eV  	\\ 
 $J$ & $0.3-0.6$ eV   	\\ 
 $t$ & $0.2-0.3$ eV  	\\ 
 \hline
\end{tabular}
\caption{
Typical estimates for model parameters relevant to iron-based superconductors \cite{Nakamura_2008, Miyake_2010}. Due to crystal field splitting, the bandwidths and chemical potentials of individual orbitals can vary significantly \cite{Yin2011}. 
The parameter $t$ refers specifically to nearest neighbor hopping.}
 \label{t1}
\end{table}

\subsection{Excitations}
The table \ref{Vc} provides contact interaction for some elementary excitations above the ground state, derived for $H_K[P=0]$.  
For realistic parameters, orbitons are strongly bound, while bi-polarons see a slightly weaker repulsion. The bi-orbiton contact energy is relative to four unbound carriers. 
The contact interaction of these objects can be reproduced from a simple two-body term of the form 
\bea
V_{pp}=V_{hh}=U'-J,\;\;\;
V_{ph}=-U'
\label{Ve}
\eea
where the subscript denotes particle-particle, hole-hole and particle-hole interaction respectively. 

\begin{table}[h!]
\centering	
\begin{tabular}{ cccc} 
 \hline
 Object & Creator & $V_{\text{contact}}$ &	Interaction	\\
 Orbiton & $a_\alpha^\dagger a_\beta$  & $-U'$ & Attractive   	\\ 
 Bi-polaron  & $a_\alpha^\dagger a_\beta^\dagger$  & $(U'-J)$ & Repulsive   	\\ 
 Bi-orbiton & $a_\alpha^\dagger a_\beta c_\gamma^\dagger c_\delta$  & $-2U'-2J$ & Attractive   	\\ 
  \hline
\end{tabular}
\caption{
Contact interaction for some common excitations above ground state on a half filled background (i.e. one electron per orbital and maximal spin), obtained for $H_K[P=0]$ . The contact energy for the bi-orbiton is relative to four unbound carriers--the energy barrier against decay into two separate orbitons is $2J$. 
}
 \label{Vc}
\end{table}

Finally, we consider the impact of the pair-conversion term in (\ref{HK}), which ``inverts'' orbitons. If we account for the contribution from single charge carriers, the energy of an orbiton is 
\bea
E_{o,\alpha\beta}=U_{\text{eff}}-U'+\delta\mu_{\alpha\beta}=J(N+1)+\delta\mu_{\alpha\beta},
\eea
where $\delta\mu_{\alpha\beta} =\mu_\beta-\mu_\alpha$ is the difference in chemical potential between the particle types . For this object to be thermodynamically stable, we require that $\delta\mu_{\alpha\beta}\sim -J(N+1)$. Applying pair conversion, we instead obtain an object with energy 

\bea
E_{o,\beta\alpha}\sim E_{o,\alpha\beta}+ 2J(N+1).
\eea
This represents a high energy state, with a correspondingly short lifetime. The primary decay path is via another pair conversion. For physically relevant model parameters, we have $P=J$, implying a slight hybridization of the two orbitons given by
\bea
|o_{GS}\rangle\sim \sqrt{1-\frac{1}{(2N+1)^2}}|o_{\alpha\beta}\rangle+\frac{1}{2(N+1)}|o_{\beta\alpha}\rangle.
\eea
The reduction of the first amplitude is small ($<0.01$ for $N=5$), and we conclude that the pair conversion term can safely be neglected. It should be stressed however, that this holds true for a doped Mott insulator, and not for a general metallic system.

\subsection{Gutzwiller-projection}
If $U_{\text{eff}}\gg t$, the lifetime of particle-hole pairs will be short, and it is a good approximation to treat such fluctuations as emergent processes, while applying Gutzwiller projection to the system \cite{PhysRevB.18.3453}. 
In multi-orbital systems, we have the prospect of bands being both hole-doped or particle doped, meaning that we either project out doublons or vacant sites in a given orbital. To simplify the representation we therefore conduct a spin-preserving particle-hole transformation of the bands that are particle doped, given by
\bea
a^\dagger_\sigma \Rightarrow  a_{-\sigma}, \;a_\sigma \Rightarrow  a_{-\sigma}^\dagger.\label{pht}
\eea
In the new representation, all carriers are hole-like. 
The delocalization part of the resulting model is described by projected hopping:
\bea
H_t=\sum_{\langle ij\rangle \alpha\sigma} t_\alpha \tilde{a}^\dagger_{i\alpha\sigma}\tilde{a}_{j\alpha\sigma}\label{Ht}
\eea
where $\tilde{a}_{i\alpha\sigma}=a_{i\alpha\sigma}(1-n_{i\alpha,-\sigma})$. It should be noted that $t_\alpha\to -t_\alpha$ under the mapping (\ref{pht}), though this only matters on non-bipartite lattices.

From the charge fluctuations, we obtain super-exchange processes of the form:
\bea\nonumber
H_S\!=\!\sum_{\langle ij\rangle,\alpha} I_\alpha 
\Bigg\{ \frac{\hat{S}_{\alpha-}^i\hat{S}_{\alpha+}^j\!+\!\hat{S}_{\alpha+}^i\hat{S}_{\alpha-}^j}{2}\!+\! \hat{S}_{\alpha z}^i\hat{S}_{\alpha z}^j\\
-\frac{1}{4}n_{i\alpha}n_{j\alpha}\Bigg\},\;I_\alpha\!=\!\frac{4 t_\alpha^2}{U_\text{eff}},
\label{HS}
\eea
where $\hat{S}^i_{\alpha\pm}$ raises/lowers the spin of the $\alpha-$orbital on site $i$ \cite{PhysRevB.18.3453}. Using Eq. (\ref{Ue}) and the estimates from the Table. (\ref{t1}), we find that $I_\alpha=0.025-0.1$ eV.

Interactions between carriers take the form 
\bea
H_V=\sum_i V_{\alpha\beta}n_{i\alpha} n_{i\beta},\label{Hv}
\eea
where $V_{\alpha\beta}$ is given by Eq. (\ref{Ve}). Note that this interaction will depend on whether the carriers were particle or hole-like in the original description. 

To take into account the Hund's coupling, we follow a strategy similar to Gutzwiller projection, and work in the subspace dictated by the projector $P^H$, (defined in Eq. \ref{PH}), while treating excitations across the Hund's gap as emergent processes.

\subsection{Hund-projected magnetism}
 Operating with $H_S$ on a ground state of the type (\ref{GS}), we obtain two principal contributions:
\bea\nonumber 
H_S|S_{M}^n,S_Z\rangle\!=\!P^HH_S|S_{M}^n,S_Z\rangle \!+\!(1\!-\!P^H)H_S|S_{M}^n,S_Z\rangle.
\eea
The latter term violates Hund's first rule to an energy cost of at least $\Delta_H$. In Matsubara theory, this exited state will decay with an imaginary lifetime of  $\sim 1/\Delta_H$, leading to emergent processes with an effective energy scale 
\bea
\epsilon\sim \frac{I^2}{4\Delta_H} \sim 0.002-0.017 \;I
\eea
near half filing. Virtual magnetic interactions of this type are thus sub-leading by $2-3$ orders of magnitude, and it is therefore an excellent approximation to discard them at temperatures $T\ll \Delta_H$ . In what follows, we will thus project out high-energy contributions from the magnetic interactions by taking
\bea
H_S\to P^H H_S.\label{HSPD}
\eea
For the spin raising/lowering operator of the electron $\alpha$, we note that $\hat{S}_{\alpha\pm}$ must necessarily alter $S_z$ by unity. Therefore, we have 
\bea\nonumber
P^H \hat{S}_{\alpha\pm}  |S_M^n,S_z\rangle &=&|S_M^n,S_z\pm 1\rangle\langle S_M^n,S_z\pm 1|\hat{S}_{\alpha\pm}  |S_M^n,S_z\rangle\\
&\equiv & |S_M^n,S_z\pm 1\rangle \gamma_{S_M^nS_z}^{\pm}.\label{SPM}
\eea
Evaluation of the inner product--see appendix I--gives
\bea\nonumber
\gamma_{S_M^nS_z}^{-}=  \frac{\sqrt{(S_M^n+S_z)(S_M^n-S_z+1)}}{2S_M^n},\\
\gamma_{S_M^nS_z}^{+}=  \frac{\sqrt{(S_M^n+S_z+1)(S_M^n-S_z)}}{2S_M^n}.
\eea

For the $\hat{S}_z$ terms in the model (\ref{HS}), we first note that these preserve $S_z$. Therefore, we must have  
\bea\nonumber
P^H \hat{S}_{\alpha z} |S_M^n,S_z\rangle\!=\!P^H \! \Big\{  u|S_M^n,S_z\rangle\!+\!v|S\!\not=\!S_M^n,S_z\rangle    \Big\}\;\;\\
=|S_M^n,S_z\rangle \langle S_M^n,S_z |\hat{S}_{\alpha z} |S_M^n,S_z\rangle=\frac{S_z}{n}|S_M^n,S_z\rangle.\;\;\label{Sz}
\eea
Combining the results (\ref{SPM}) and (\ref{Sz}), we can write the projected  Hamiltonian (\ref{HSPD}) as 
\bea\nonumber
P^H H_S=I^T_{ij}\sum_{\langle ij\rangle} \Bigg\{\frac{\hat{S}_{-}^i\hat{S}_{+}^j+\hat{S}_{-}^j\hat{S}_+^i}{2 n_i n_j} + \frac{\hat{S}_z^i}{n_i}\frac{\hat{S}_z^j}{n_j} \Bigg\}
\\
 -\frac{1}{4} \sum_\alpha I_\alpha n_{i\alpha}n_{j\alpha},
\;
I^T_{ij}=\sum_\alpha  I_\alpha n_{i\alpha}n_{j\alpha}.\;  \label{HSP}
\eea
At half filing, the spin sector is thus governed by an emergent spin-$N/2$ Heisenberg model. For $N=5$, mean-field theory--see Appendix II--suggests that quantum spin fluctuations are modest at approximately $4\%$ on the square lattice, and even smaller in three dimensions. This places us close to the classical limit, with a coupling strength 
\bea
J_\text{classical}=\frac{1}{4}\sum_\alpha I_\alpha \sim 0.03 - 0.125 \text{ eV}.
\eea
On a cubic lattice, this would give an ordering temperature of $T_c\sim 500-2100 \text{K}$, though it should be stressed that many Hunds metals feature a weaker inter-layer interaction and that the presence of carriers tends to reduce the ordering temperature.

\subsection{Charge carriers}
To establish the role of mobile dopants, we construct the analogue of the delocalization part of the Hamiltonian [Eq. (\ref{Ht})] within the Hund-projected subspace.
We begin by defining the normalization coefficients of the spin ladder operators as follows:

\bea
\xi_{S,S_z}^\pm \hat{S}_\pm|S,S_z\rangle =|S,S_z\pm1\rangle,\;|S_z\pm1|\le S\\
\chi_{S,l}^\pm \hat{S}_\pm^l |S,\mp S\rangle =|S,\pm(l-S)\rangle,\;|S-l|\le S .
\eea
Consider a site with $n$ particles in a state that maximizes total spin. If the orbital $\alpha$ is vacant, then it can be filled by applying the corresponding creation operator. 
Assuming $S_z=-S_M^{n}+l$ in the initial state we find
\bea \nonumber
a_{\uparrow\alpha}^\dagger|S_M^{n},-S_M^{n}+l\rangle
=a_{\uparrow\alpha}^\dagger \hat{S}_+^l|S_M^{n},-S_M^{n}\rangle \chi^+_{S^{n}_M,l} \\
=  \hat{S}_+^l \hat{S}_{\alpha +}|S_M^{n+1},-S_M^{n+1}\rangle \chi^+_{S^{n}_M,l},\label{create}
\eea 
where we have used $[\hat{S}_+^l, c_{\uparrow\alpha}^\dagger]=0$ and $a^\dagger_{\uparrow\alpha}\to \hat{S}_{\alpha +} \hat{a}^\dagger_{\downarrow\alpha}$. The operator $\hat{S}_{\alpha+}$ splits the initial state into two contributions,
\bea\nonumber
\hat{S}_{\alpha+}|S_M^{n+1},-S_M^{n+1}\rangle=\frac{1}{\sqrt{n+1}}|S_M^{n+1},1\!-\!S_M^{n+1}\rangle\!+|R_\alpha\rangle\;\;\\\label{Ra}
\eea
where only the former lies in the Hund subspace. 
Under the subsequent spin-raising operations, the first part transforms as follows (see Appendix III):
\bea\nonumber
\hat{S}_+^l|S_M^{n+1},1\!-\!S_M^{n+1}\rangle  \frac{\chi^+_{S^{n},l}}{\sqrt{n+1}}
=|S_M^{n+1},l\!+\!1\!-\!S_M^{n+1}\rangle \sqrt{\frac{l\!+\!1}{n+1}}.\\ \label{createII}
\eea
The second contribution takes the form
\bea
|R_\alpha\rangle = |S_M^{n+1}-1,1-S_M^{n+1}\rangle_\alpha
\eea
which is an eigenstate of $H_K$ with energy $\Delta_H$ (even after applying $\hat{S}_+^l$). Since $\Delta_H\gg t,j$, this state must have a short lifetime, giving rise to emergent processes with energy scales 
\bea
\frac{t I_\beta}{2\Delta_H\sqrt{N}}\sim 0.002-0.015 t,\\
\frac{t^2 }{\Delta_H}\sim 0.07-0.2 t, \label{kineticDecay}
\eea
depending on whether it decays by magnetic or kinetic mechanisms. The former is sub-leading by 2-3 orders of magnitude, and may safely be neglected. 
The kinetic decay is an order of magnitude smaller than the hopping integral, but has the ability to nucleate a weak pair-propagation mechanism under certain circumstances--see discussion below on ``Excitations across the Hund's gap''.

In the next stage, we consider the creation of a hole on a state with a filled $\alpha-$orbital. Assuming $S_z=S_M^n-m$, we find
\bea\nonumber
a_{\uparrow\alpha}|S_M^n,S_M^n-m\rangle=a_{\uparrow\alpha} \hat{S}_-^m |S_M^n, S_M^n\rangle \chi^-_{S_M^n,m}\;\;\;\;\\
=\!\hat{S}_-^m |S_M^{n-1}\!,S_M^{n-1}\rangle \chi^-_{S_M^n,m}\!=\!|S_M^{n\!-\!1},S_M^{n\!-\!1}\!\!-\!m\rangle \frac{\chi^-_{S_M^n,m}}{\chi^-_{S_M^{n-1},m}}\;\;\;\;
\eea
where we have used $[c_{\uparrow,\alpha}, \hat{S}_-]=0$. Next we note that
\bea
\frac{\chi^-_{S_n^n,m}}{\chi^-_{S_M^{n-1},m}}=\sqrt{\frac{n-m}{n}}\equiv \sqrt{\frac{l}{n}},\label{annihilation} 
\eea
since $m$ counts the number of spin flips from $S_z=S_M^n$, while $l$ counts the number of flips from $S_z=-S_M^n$. 
Combining (\ref{createII}) and (\ref{annihilation}) we finally conclude
\bea\nonumber
P^H a^\dagger_{\uparrow\alpha}|S_M^{n}, l\!-\!S_M^n\rangle\!=\!|S_M^{n+1}, l\!+\!1\!-\!S_M^{n+1}\rangle   \sqrt{\frac{l\!+\!1}{n\!+\!1}}\\
 a_{\uparrow\alpha}|S_M^{n},l-S_M^{n}\rangle=|S_M^{n-1},l-1-S_m^{n-1}\rangle \sqrt{\frac{l}{n}},\label{caup}
\eea
where it is assumed that the $\alpha-$orbital of the initial state is vacant or occupied respectively. 
In the Hund-projected subspace, a state is uniquely identified by the orbital occupation $\bar{o}$ and the spin $S_z$. Hence, we adapt the notation
\bea
|\bar{o},l\rangle,\; \text{with}\;\;0 \le l\le n[\bar{o}],
\eea
where $n$ is obtained from Eq. (\ref{defNS}).
The spin is then mapped onto a bosonic field so that 
\bea
\hat{l}= \hat{b}^\dagger \hat{b}
\eea
counts the number of spin up electrons. 
The orbital occupation is encoded into $N$ fields of spin-less fermions in such a way that the charge-carrying holes are mapped onto particles rather than hole. This gives 
\bea
\hat{n}=N-\sum_\alpha c^\dagger_\alpha c_\alpha,
\eea 
where $c_\alpha^\dagger,\;c_\alpha$ are creation/annihilation operators of a spin-less fermion. 
The original creation and annihilation operators can then be written
\bea\nonumber
P^H \hat{a}^\dagger_{\uparrow\alpha}P^H= \hat{c}_\alpha \hat{b}^\dagger\frac{1}{\sqrt{\hat{n}+1}} ,&&\;\;\;\;\hat{a}_{\uparrow\alpha}= \hat{b} \hat{c}^\dagger_\alpha \frac{1}{\sqrt{\hat{n}}}\\
P^H \hat{a}^\dagger_{\downarrow\alpha}P^H= \hat{c}_\alpha\sqrt{\frac{\hat{n}+1-\hat{l}}{\hat{n}+1}} && ,\;\;\; \hat{a}_{\downarrow\alpha}= \hat{c}^\dagger_\alpha \sqrt{\frac{\hat{n}-\hat{l}}{\hat{n}} }.\label{OperatorTransform}
\eea
Note that the bosonic operators $\hat{b}^\dagger,\hat{b}$ have matrix elements $\sim \sqrt{\hat{l}+1}$ and $\sqrt{\hat{l}}$ respectively to reproduce Eq. (\ref{caup}).
This gives
\bea
P^H H_t P^H=-\sum_{\langle ij\rangle \alpha} t_\alpha( c^\dagger_{i\alpha} c_{j\alpha} \hat{b}_{i}\hat{b}^\dagger_{j}\hat{\gamma}^{\uparrow}_{ij}
+c^\dagger_{i\alpha}  c_{j\alpha}\hat{\gamma}_{ij}^\downarrow),\;\;\;\label{lowEhopping}
\eea
where the sign results from commuting $c c^\dagger\to -c^\dagger c$, and
\bea\nonumber
\hat{\gamma}_{ij}^\uparrow =\frac{1}{\sqrt{\hat{n}_i}}\frac{1}{\sqrt{\hat{n}_j+1}},\;\;
\hat{\gamma}_{ij}^\downarrow = \sqrt{1-\frac{\hat{b}_i^\dagger \hat{b}_i}{\hat{n}_i}}\sqrt{1-\frac{\hat{b}_j^\dagger \hat{b}_j}{\hat{n}_j+1}}.\\
\label{gamma}
\eea

\subsection{Hund-projected model}
Combining the hopping integral (\ref{lowEhopping}), the inter-carrier interaction (\ref{Hv}) and the super-exchange (\ref{HSP}) we obtain the Hund-projected Kanamori model:
 
\bea\nonumber
H=\sum_{\langle ij\rangle \alpha} \tilde{t}_\alpha c^\dagger_{i\alpha} c_{j\alpha} \hat{b}_{i}\hat{b}^\dagger_{j}\hat{\gamma}^{\uparrow}_{ij}
+c^\dagger_{i\alpha}  c_{j\alpha}\hat{\gamma}_{ij}^\downarrow+\sum_i V_{\alpha\beta}n_{i\alpha} n_{i\beta},\\
+I^T_{ij}\sum_{\langle ij\rangle} \Bigg\{\frac{\hat{S}_{-}^i\hat{S}_{+}^j+\hat{S}_{-}^j\hat{S}_+^i}{2 n_i n_j} + \frac{\hat{S}_z^i}{n_i}\frac{\hat{S}_z^j}{n_j} \Bigg\}
 -\frac{1}{4} \sum_\alpha I_\alpha n_{i\alpha}n_{j\alpha},
 \nonumber\\
 \label{effectiveModel}
\eea
where $\tilde{t}_\alpha=\pm t_\alpha$ and the sign depends on whether the orbital $\alpha$ was particle like ($+$) or hole-like ($-$) in the original representation. 
The spin-less fermions in this model represent charge carriers, while the bosonic occupation number counts the number of spin-up electrons on a site. The latter is constrained to 
\bea
\hat{l}= \hat{b}^\dagger \hat{b}\le \hat{n}=N-\sum_\alpha c^\dagger_\alpha c_\alpha.
\eea
The interactions in the model (\ref{effectiveModel}) are given by
\bea
V_{\alpha\beta}= U'-J \;\;  \text{or}
 =-U',
\eea
for same and different charges respectively. 
The effective hopping integral is
\bea\nonumber
\hat{\gamma}_{ij}^\uparrow =\frac{1}{\sqrt{\hat{n}_i}}\frac{1}{\sqrt{\hat{n}_j+1}},\;\;
\hat{\gamma}_{ij}^\downarrow = \sqrt{1-\frac{\hat{b}_i^\dagger \hat{b}_i}{\hat{n}_i}}\sqrt{1-\frac{\hat{b}_j^\dagger \hat{b}_j}{\hat{n}_j+1}}.
\eea
and the super-exchange term takes the form
\bea
I^T_{ij}=\sum_\alpha  I_\alpha  c_{i\alpha}c_{i\alpha}^\dagger c_{j\alpha}c_{j\alpha}^\dagger. 
\eea
For a low carrier density, we can make the approximation 
\bea
\hat{\gamma}^\uparrow\approx\frac{1}{N},\;\;\;
\hat{\gamma}^\downarrow_{ij}\approx \sqrt{1-\frac{\hat{b}_i^\dagger \hat{b}_i}{N}}\sqrt{1-\frac{\hat{b}_j^\dagger \hat{b}_j}{N}}\label{lowP}
\eea
which becomes exact for isolated dopants. In this scenario we can also use
\bea
I^T_{ij}\approx \sum_\alpha  I_\alpha, 
\eea 
which becomes exact in the absence of carriers.

\subsection{Excitations across the Hund's gap}
To complete our discussion of dopants, we now turn to kinetic processes that transiently excite the system across the Hund's gap (see Eq. \ref{kineticDecay}). Such processes involve intermediate states in which the local spin configuration temporarily violates Hund's rule, incurring an energy cost on the order of $\Delta_H$. Starting from Eqs. ({\ref{create}-\ref{Ra}), the corresponding excited state can be expressed as

\bea\nonumber
\chi^+_{S_M^n,l}\hat{S}_+^l |R_\alpha\rangle\equiv \lambda_{n,l}\chi^+_{S_M^n,l}\hat{S}_+^l \Bigg|\frac{n}{2}-1,1-\frac{n}{2}\Bigg\rangle_{\alpha},
\eea
where $\lambda$ is the amplitude of the state.
The subscript $\alpha$ here indicates that the excited state was generated by creating a particle in the $\alpha-$orbital. This is important since states which do not maximize total spin are typically degenerate so that $|R_\alpha\rangle\not=|R_\beta\rangle$ in general. 

Applying an annihilation operator and projecting on the Hund's ground state we find 
\bea
 P^H a_{\beta\sigma} \lambda_{n,l}\chi^+_{S_M^n,l}\hat{S}_+^l \Bigg|\frac{n}{2}-1,1-\frac{n}{2}\Bigg\rangle_\alpha \\
\equiv \kappa_{\alpha \beta \sigma l}\Bigg|\frac{n-1}{2},l+1-\frac{n+\sigma}{2}\Bigg\rangle_{o'}, 
\eea
where $o'$ denotes the orbital occupation after annihilating a particle in the band $\beta$, and $\kappa$ is the amplitude of the resulting state.

To derive the emergent process associated with kinetic excitation and decay, consider two subsequent tunneling events:
\bea
 t^2 a^\dagger_{i\alpha\sigma}(\tau)a_{j\alpha\sigma}(\tau) a^\dagger_{j\beta\sigma'}(\tau')a_{k\beta \sigma'}(\tau'),\;\tau>\tau'.
\eea
Here, we are interested in the contribution when  $\alpha^\dagger_{j\beta \sigma'}$ excites the site $j$ across the Hund's gap. The lifetime of this state is $\delta\tau= 1/\Delta_H$ and with $\sigma'=\;\uparrow$, we obtain an emergent pair propagation term
\bea\nonumber
H_h\!\sim\!\frac{-t^2}{\Delta_H}   c_{j\alpha}^\dagger c_{j\beta}  \Bigg(\delta_{\sigma,\uparrow}\!+\!\delta_{\sigma,\downarrow}\frac{\hat{b}_j^\dagger}{\sqrt{\hat{l}_j+1}}\Bigg)  \kappa_{\alpha\beta \sigma l_j}  P^H a^\dagger_{i\alpha\sigma}  a_{k\beta \uparrow},\\
\label{pairProp}
\eea
where $a^\dagger,a$ take the form (\ref{OperatorTransform}). Some values of $\kappa$ are given in table \ref{kappa}. These indicate that pair propagation (\ref{pairProp}) is strongly dependent on the spin background, and notably vanishes for a polarized system. 

The ideal circumstances for pair propagation is on an antiferromagnetic background. 
Assuming that $S_{z}=S_M^n$ on $i,k$, $S_{z}=-S_M^n$ on $j$ and $\sigma=\sigma'=\uparrow$ we find
\bea
H_h\sim -t_p   c_{j\alpha}^\dagger c_{j\alpha}  \hat{c}_{i\alpha}\frac{ \hat{b}_i^\dagger}{\sqrt{\hat{n}_i+1}}   c^\dagger_{k\alpha}    \frac{\hat{b}_k}{\sqrt{\hat{n}_k}}\\
\text{with}\;t_p=\frac{t^2}{\Delta_H}\kappa_{AF},\;\;\;\kappa_{AF}=1-\frac{1}{n}.
\eea
Here, we obtain two principal contributions corresponding to $i=k$ and $i\not=k$. Assuming that the sites are not occupied by other carriers, we can use (\ref{lowP}). For $i=k$ we find 
\bea
H_{NN}= -t_p\sum_{\langle ij\rangle\alpha}  c_{j\alpha}^\dagger c_{j\alpha}  \hat{c}_{i\alpha}   c^\dagger_{i\alpha}    
\eea
which describes a nearest neighbor repulsion with magnitude $t_p$. For $i\not=k$ we find 
\bea
H_p=-t_p\sum_{\langle ij\rangle\langle jk\rangle\alpha}
     c^\dagger_{i\alpha}  c_{j\alpha}   c_{j\alpha}^\dagger   \hat{c}_{k\alpha}  \frac{ \hat{b}_k^\dagger \hat{b}_i}{N}  
\eea
which describes pair propagation. For opposite spins (i.e. $S_z=-S_M^n$ on $i,k$ etc) we analogously find  
\bea
H_p=-t_p\sum_{\langle ij\rangle\langle jk\rangle\alpha}
     c^\dagger_{i\alpha}  c_{j\alpha}   c_{j\alpha}^\dagger   \hat{c}_{k\alpha}   
\eea
by propagating spin down electrons. 
We thus obtain two principal contributions from transient excitations across the Hund's gap in an antiferromagnet: Nearest neighbor repulsion and a form of pair propagation that preserves the magnetic order. The potency of this mechanism diminishes as antiferromagnetic correlations weaken, and vanishes completely for a ferromagnet.

\begin{table}[h!]
\centering	
\begin{tabular}{ cccccc} 
 \hline
 Operator & $l=0$ & $l=1$ & $l=2$ &	$l=3$ & $l=4$	\\
 $a_{\alpha\uparrow}$ & 0.8 & 0.54 & 0.31 & 0.13 & 0 \\
 $a_{\alpha\downarrow}$ & 0.4 & 0.44 & 0.38 & 0.26 & 0\\ 
 $a_{\beta\not=\alpha\uparrow}$ & 0.2 & 0.13 & 0.08 & 0.03 & 0\\
 $a_{\beta\not=\alpha\downarrow}$ & 0.1 & 0.11 & 0.095 & 0.063 & 0\\ 
  \hline
\end{tabular}
\caption{
Numerical values of $\kappa$ obtained for $n=N=5$, i.e. a state $|R_\alpha\rangle$ with five electrons in five orbitals. 
}
 \label{kappa}
\end{table}

Excitations across the Hund's gap can be treated systematically in path integral Monte Carlo simulations by introducing $N$ additional bosonic fields to describe states $|R_\alpha\rangle$. The parameters $\lambda$ and $\kappa$ will have to be computed numerically and additional kinetic processes introduced to couple the Hund-projected subspace to excited states.

\subsection{Discussion}

In conclusion, we have derived an effective low-energy model of a Hund's metal near the Mott insulating regime. The undoped system exhibits strong antiferromagnetism. For five bands, quantum spin fluctuations are found to be small, placing the system close to the classical limit (see Appendix II).
The motion of doped carriers depends strongly on the local spin correlations (see Eq. \ref{gamma}), driving a form of kinetic ferromagnetism that is notably different from other described cases: It does not depend on the interference of paths, but instead emerges at the level of effective hopping. This points to a mechanism far more potent than that of conventional single-band lattice models, which are frequently relatively weak \cite{PhysRevB.2.1324}. The emergence of pronounced Nagaoka-type \cite{PhysRev.147.392} ferromagnetism is in good agreement with first-principles studies.

The original Kanamori model is notoriously challenging from a computational perspective, and most existing theoretical results rely on dynamical mean-field theory (DMFT). By contrast, the Hund-projected Kanamori Model derived here offers a far simpler and more tractable framework. Notably, the sign problem is extensive only in the number of mobile carriers, as sign changes arise solely from combined spin and charge motion. 
This property enables the use of advanced path-integral techniques-such as the Worm algorithm \cite{Prokofev1998} and suitable diagrammatic methods \cite{PhysRevB.103.195147,PhysRevB.97.075119,0953-8984-29-38-385602}-to study few-carrier problems in detail, allowing direct access to the structure of spin polarons \cite{blomquist2019ab} and the effective interactions between them \cite{PhysRevResearch.3.013272}. These observables are generally inaccessible within DMFT.
Our work thus establishes a controlled route from the microscopic Kanamori Hamiltonian to a low-energy theory suitable for high-precision numerical studies. 
The resulting model captures the essential physics of Hund's metals: the intricate interplay between spin correlations, carrier motion, and Hund's coupling that underlies their anomalous electronic structure.



This work was supported by the Swedish Research Council (VR) through grant 2018-03882, Stiftelsen Olle Enquist through grant 240-0803, and the Knut and Alice Wallenberg Foundation, project KAW 2024.0131
 The computations were enabled by resources provided by the National Academic Infrastructure for Supercomputing in Sweden (NAISS), partially funded by the Swedish Research Council through grant agreement no. 2022-06725.
The author would like to thank Luca de Medici for important discussion.

\bibliography{biblio.bib}

\section{Appendix}

\subsection{I. Matrix elements of the spin lowering/raising operator}
For the spin lowering operator of the electron $\alpha$, we note that $S_\alpha^-$ must necessarily reduce $S_z$ by unity. Therefore, we have 
\bea\nonumber
P^H \hat{S}_{\alpha}^{-}  |S_M,S_z\rangle=|S_M,S_z-1\rangle\langle S_M,S_z-1|\hat{S}_{\alpha}^-  |S_M,S_z\rangle.\\
\label{PHSM2}
\eea
The inner product can be rewritten as 
\bea\nonumber
\langle S_M,S_M|(\hat{S}^+)^{l+1} \hat{S}_{\alpha}^- (\hat{S}^-)^{l}|S_M,S_M\rangle \Pi_{a=0}^{l}\xi_a \Pi_{b=0}^{l-1}\xi_b  \\
\eea
where 
\bea
\xi_a^{-1}=\sqrt{(S_M+[S_M-a])(S_M-[S_M-a]+1)} \label{matEl}
\eea
are the coefficients for the spin lowering operators with $S_Z=S_M-a$.
 Since $[\hat{S}_\alpha^-,\hat{S}^-]=0$ we can write  
\bea\nonumber
\langle S_M,S_M|(\hat{S}^+)^{l+1}  (\hat{S}^-)^{l} \hat{S}_{\alpha}^-  |S_M,S_M\rangle\Pi_{a=0}^{l}\xi_a \Pi_{b=0}^{l-1}\xi_b\\\nonumber
=\langle S_M,S_M|(\hat{S}^+)^{l+1}  (\hat{S}^-)^{l}  |S_M,S_M-1\rangle\\\nonumber
\times \langle S_M,S_M-1|  \hat{S}_{\alpha}^-  |S_M,S_M\rangle\Pi_{a=0}^{l}\xi_a \Pi_{b=0}^{l-1}\xi_b\\\nonumber
=\langle S_M,S_M|(\hat{S}^+)^{l+1}  (\hat{S}^-)^{l+1}  |S_M,S_M\rangle\\\nonumber
\times  \frac{\xi_{0}}{\sqrt{N}} \Pi_{a=0}^{l}\xi_a \Pi_{b=0}^{l-1}\xi_b=\frac{\xi_0\xi_{l}^{-1}}{\sqrt{N}}\equiv \gamma_{S_MS_z}^{-}
\eea

Using \ref{matEl} and $\xi_0=N^{-1/2}=(2S_M)^{-1/2}$ we find the inner product
\bea\nonumber
\gamma_{S_MS_z}^{-}=  \frac{\sqrt{(S_M+S_z)(S_M-S_z+1)}}{2S_M},\\
\gamma_{S_MS_z}^{+}=  \frac{\sqrt{(S_M+S_z+1)(S_M-S_z)}}{2S_M}\label{innerProd2}
\eea
where the latter follows by symmetry. 
 Combining (\ref{PHSM2}) and (\ref{innerProd2}) we conclude
\bea
P^H \hat{S}_\alpha^{\pm} |S_{M},S_Z\rangle = \gamma_{S_MS_z}^{\pm}\hat{S}_N^\pm |S_{M},S_Z\rangle,
\label{SPM2}
\eea 
where $\hat{S}_N^\pm|S,S_z\rangle=|S,S_z\pm 1 \rangle$ for $|S_z\pm 1|\le S$ defines the normalized spin raising/lowering operator.

\subsection{II. Holstein-Primakoff mean field theory}
As a starting point, we consider the Hund's projected spin Hamiltonian at half filling
\bea\nonumber
P^H H_S=I_T\sum_{\langle ij\rangle} \Bigg\{\frac{S_{i}^-S_{j}^++S_{j}^-S_i^+}{2N^2} + \frac{S_i^z}{N}\frac{S_j^z}{N} \Bigg\}
\label{APP-HSP}
\eea
In the next stage, we apply a Holstein-Primakoff type map of the form
\bea\nonumber
S_i^z=S-a_i^\dagger a_i,\;\;S_{i}^+=a_i\sqrt{N},\;\; S_{i}^-=a^\dagger_i\sqrt{N}\;\;\;\;\text{A-lattice}\\
S_j^z=b_j^\dagger b_j-S,\;\;S_{j}^+=b_j^+\sqrt{N},\;\;S_{j}^-=b_j\sqrt{N}\;\;\;\;\text{B-lattice}\label{APP-Holstein-Primakov}\nonumber\\
\eea
Where $S^\dagger=a_i\sqrt{N}$ etc result from the approximation $\langle a_i^\dagger a_i\rangle$ and $\langle b_i^\dagger b_i\rangle\ll N$. 
We find 
\bea\nonumber
P^H H_S&=&I_T\sum_{\langle ij\rangle} \Bigg\{\frac{a_i^\dagger b_j^\dagger+a_i b_j}{2N} \\&+& \frac{-S^2 -a_i^\dagger a_i b_j^\dagger b_j+S(a_i^\dagger a_i+b_j^\dagger b_j)}{N^2} \Bigg\}.
\label{APP-HSP}
\eea
The quartic term is the smallest, and scales as $\sim A_s^2/N$, where $A_s$ denotes the spin-wave amplitude, so we proceed to drop it. 

Noting that $S=N/2$ we find   
\bea\nonumber
P^H H_S\approx I_T\sum_{\langle ij\rangle} \Bigg\{\frac{a_i^\dagger b_j^\dagger+ a_ib_j+a_i^\dagger a_i+b_j^\dagger b_j}{2N}-\frac{1}{4}\Bigg\},
\eea
where the last term is the classical energy. 
Fourier transforming, the quantum part becomes
\bea\nonumber
H_\text{Q}=\sum_\bk A_\bk(a^\dagger_\bk a_k+b^\dagger_\bk b_\bk)+B_\bk (a_\bk b_{-\bk}+a_\bk^\dagger b_{-\bk}^\dagger) \\\nonumber
A_k=\frac{zI_T}{2N},\;B_\bk=\frac{zI_T}{2N}\gamma_\bk,\;\gamma_\bk=\frac{1}{z}\sum_\delta e^{i\bk\cdot \delta}\\
\eea

Adopt a spinor notation $\Psi_\bk^\dagger=[a^\dagger_\bk,b_{-\bk}],\; \Psi_\bk=[a_\bk,b_{-\bk}^\dagger]$ to obtain
\bea
H_Q=\sum_k \Psi_\bk^\dagger h_\bk \Psi_\bk,\;h_\bk=(A_\bk\sigma_0+B_\bk\sigma_x).
\eea
The time evolution of an operator $\Psi_\bk$ is given by
\bea
i\partial_t \Psi_\bk=[\Psi_\bk,H_\text{Q}]=\Lagr_\bk\Psi_\bk,\;\Lagr_\bk=\sigma_3 h_\bk.
\eea
The appropriate Bogolyubov transform is 
\begin{equation*}
T = 
\begin{bmatrix}
\cosh r &-\sinh r\\
-\sinh r& \cosh r
\end{bmatrix},\; r=\frac{1}{2}\arctan \frac{B}{A}.
\end{equation*}

This gives
\bea
T^{-1} \Lagr_\bk T =\omega_\bk \sigma_3=D_\bk,\;\omega_\bk=\sqrt{A^2_\bk-B_\bk^2}.
\eea
Introducing new operators 

\begin{equation*}
\Psi_\bk=T\Phi_\bk,\; \Phi_\bk =
\begin{bmatrix}
\alpha_\bk \\
\beta_{-\bk}^\dagger  
\end{bmatrix},
\end{equation*}
the time evolution is given by 
\bea
i\partial_t \Phi_\bk=D_\bk\Phi_\bk
\eea
implying a Hamiltonian matrix
\bea
\tilde{k}_\bk=\sigma_3 D_\bk=\omega_\bk\sigma_0.
\eea
In the ground state, we find 
\bea
\alpha_\bk|0\rangle=\beta_\bk|0\rangle=0,
\eea
implying
\bea
\langle a_\bk^\dagger a_\bk\rangle=v_\bk^2=\frac{1}{2}\Bigg( \frac{A_\bk}{\sqrt{A_\bk^2-B_\bk^2}} -1\Bigg).
\eea
The total fluctuation is then given by
\bea
\langle a_i^\dagger a_i\rangle=\sum_k \langle a_\bk^\dagger a_\bk\rangle=\sum_\bk \frac{1}{2}\Bigg( \frac{1}{\sqrt{1-\gamma_\bk^2}} -1\Bigg).
\eea
On a square/cubic lattice we find 
\bea
\gamma_\bk=\frac{1}{2D}\sum_{l=1}^{D}2\cos k_l
\eea
where $D$ is the dimension. On the square lattice, we obtain

\bea
\langle a_\bk^\dagger a_\bk\rangle\approx 0.197.
\eea
This result is independent of the total spin. For a five band model, we thus find that fluctuations are approximately $4\%$ in a given band.

\subsection{III. Spin rotation of Hund's ground state}

Under the subsequent spin-raising operations, the first part of (\ref{Ra}) transforms as follows:
\bea\nonumber
\hat{S}_+^l|S_M^{n+1},1-S_M^{n+1}\rangle \frac{\chi^+_{S^{n}_M,l}}{\sqrt{n+1}}\\\nonumber
=\hat{S}_+^l \hat{S}_+|S_M^{n+1},-S_M^{n+1}\rangle \frac{\chi^+_{n,l}}{\sqrt{n+1}} \xi^+_{S_M^{n+1},-S_M^{n+1}}\\\nonumber
= |S_M^{n+1},l+1-S_M^{n+1}\rangle \frac{\chi^+_{n,l}}{\chi^+_{S_M^{n+1},l+1}\sqrt{n+1}} \xi^+_{S_M^{n+1},-S_M^{n+1}}\\\nonumber
=|S_M^{n+1},l+1-S_M^{n+1}\rangle \sqrt{\frac{(l+1)}{n+1}}\\
\eea
where $l+1=S_M^N+S_z+\frac{1}{2}$.

\

\end{document}